\documentclass[%
reprint,
superscriptaddress,
twocolumn,
 amsmath,amssymb,
 aps,
 prl,
]{revtex4-1}

\usepackage{graphicx}
\usepackage{dcolumn}
\usepackage{bm}
\usepackage{amsmath}
\usepackage{epsfig}
\usepackage{rotating}
\usepackage{enumerate}
\usepackage{xcolor}

\begin{document}

\title{The ferroelectric domain wall phonon polarizer}

\author{Miquel Royo}
\email{mroyo@icmab.es} 
\affiliation{Institut de Ci\`encia de Materials de Barcelona (ICMAB--CSIC)
             Campus de Bellaterra, 08193 Bellaterra, Barcelona, Spain}

\author{Carlos Escorihuela-Sayalero}
\affiliation{Materials Research and Technology Department,
             Luxembourg Institute of Science and Technology,
             41 rue du Brill, L-4422 Belvaux, Luxembourg}

\author{Jorge \'I\~{n}iguez}
\email{jorge.iniguez@list.lu}
\affiliation{Materials Research and Technology Department,
             Luxembourg Institute of Science and Technology,
             41 rue du Brill, L-4422 Belvaux, Luxembourg}

\author{Riccardo Rurali}
\email{rrurali@icmab.es}
\affiliation{Institut de Ci\`encia de Materials de Barcelona (ICMAB--CSIC)
             Campus de Bellaterra, 08193 Bellaterra, Barcelona, Spain}

\date{\today}

\begin{abstract}
Modulating the polarization of a beam of quantum particles is a powerful method to tailor the macroscopic 
properties of the ensuing energy flux as it directly influences the way in which its quantum constituents 
interact with other particles, waves or continuum media. 
Practical polarizers, being well developed for electric and electromagnetic energy, have not been proposed to date
for heat fluxes carried by phonons.  
Here we report on atomistic phonon transport calculations demonstrating that ferroelectric domain walls can operate as phonon polarizers
when a heat flux pierces them. Our simulations for representative ferroelectric perovskite PbTiO$_3$ show that the structural 
inhomogeneity associated to the domain walls strongly suppresses transverse phonons, while longitudinally polarized modes can travel through 
multiple walls in series largely ignoring their presence. 
\end{abstract}

\maketitle

Controlling heat conduction has always been a challenging task, to the point that our ability to manipulate thermal fluxes lags 
far behind our long-standing know-how in manipulating electric and electromagnetic fluxes. In solids, the principal complication 
in manipulating heat stems from the \textit{a priori} impossibility to use electrical signals or electromagnetic fields for that purpose. 
This is because phonons, the dominant heat carriers in insulating materials, do not possess a bare charge.

Heat conduction can be instead modulated following basically two fundamentally different approaches. In the 
first approach, the most extended one, structural inhomogeneities such as atomic-scale defects, materials interfaces or surfaces, are incorporated in a physical material to scatter phonons diffusively and reduce the thermal 
conductivity.\cite{Toberer2012,Cahill2014} In addition to this approach, which exploits the corpuscular nature of phonons, wave-interference phenomena
have been recently demonstrated to effectively block the propagation of phonons traveling through periodic superlattices 
or phononic crystals.\cite{Luckyanova2012,Ravichandran2013a,Maldovan2015} Interestingly, both approaches offer the possibility to focus on a fixed frequency window, i.e., to perform a frequency filtering effect, by choosing the size of the structural inhomogeneities or their periodicity.

However, candidate systems to effectively filter phonons of a certain polarization, the so-called phonon polarizers, are yet
to come forth. The only previous hints of a mode-dependent phonon scattering have been observed by means of ballistic phonon imaging techniques at very low temperatures 
($<3$ K) in highly dislocated LiF crystals\cite{Northrop1982} and in an exotic ferroelectric as $\mathrm{KH_2PO_4}$.\cite{Weilert1993} Nevertheless, in these systems a clear polarizer
effect capable to discern between, at least, longitudinal and transverse phonons, was not observed. It has also been
suggested that material interfaces enclosing a molecular self-assembled monolayer should exhibit a phonon-polarizing 
effect due to the anisotropy in the bond strength across the length and breadth of the monolayer, \cite{Wang2006} but such a conjecture
has not been experimentally nor numerically proved. 
The polarization of phonons influences important physical phenomena 
wherein phonons participate. For example the way in which phonons 
couple to defects and to electronic states depends markedly on their 
polarization, as well as on other defining properties such as their 
frequency and propagation direction. Likewise, phonons play an 
important role in nonradiative energy relaxation and inter-band 
recombination processes, which are known to be ruled by 
polarization-dependent selection rules. Therefore, the possibility 
to modulate the polarization of a flux of phonons at will would offers 
definite possibilities for modifying the behavior of materials and/or 
carrying information.

Here we demonstrate that the ferroelectric interfaces occurring between polarization domains in a well known perovskite oxide,
$\mathrm{PbTiO_3}$, can effectively operate as phonon polarizers. Such interfaces, also known as domain walls (DWs), are today
routinely created, moved and annihilated by application of local electric fields, \cite{Nelson2011,McGilly2015,Ievlev2014} a possibility that 
renders them potential candidates to dynamically manipulate phonons with electrical signals and 
to enable in-situ reconfigurable thermal circuiting within a given material volume. In fact, DW thermal transport engineering 
was already attempted in the last decades in ferroelastics \cite{Ding2015,Li2014a} and ferroelectrics such as $\mathrm{BaTiO_3}$ \cite{Mante1971} and 
$\mathrm{KH_2PO_4}$ \cite{Weilert1993,Weilert1993a} obtaining 
a remarkable modulation of the conductivity by means of strain and electrical alteration of the DW density in the crystals.
Today the focus is on ferroelectric thin films that permit a precise control of the spacing between DWs which, in turn, makes 
it possible to control the dominant scattering mechanism: For DW separations comparable or below the average phonon mean-free path, 
the phonon transport is dominated by the walls scattering rather than by anharmonic phonon-phonon interactions. Ihlefeld \textit{et al.} \cite{Ihlefeld2015,Hopkins2013} have recently proceeded in this way to prepare the first prototypes of electrically actuated DW thermal 
switches operating over a broad temperature range, including room temperature. 

\begin{figure}
\includegraphics[width=1.0\linewidth]{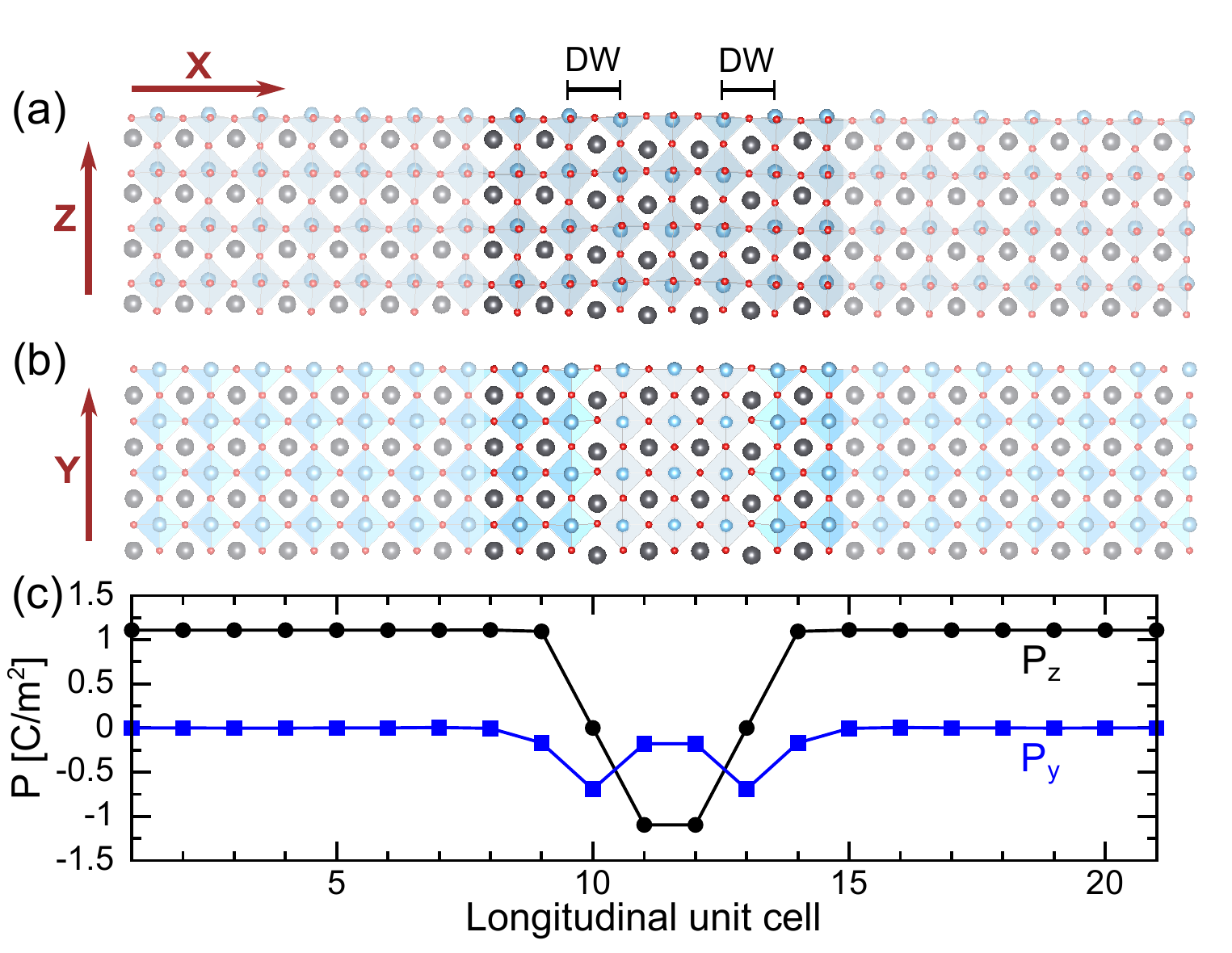}
\caption{Views of the $x,z$ (a) and $x,y$ (b) cross sections of a prototypical tetragonal PTO supercell employed in the computational simulations. Grey, blue
and red colours are used for Pb, Ti and O atoms, respectively, and partial transparency is used to illustrate atomic positions in
the homogeneous contacts. The channel includes two flat domain walls extending over a Pb-O ($y,z$) plane and separated by approximately
two primitive cells. Panel (c) shows the polarization profiles along the thermal transport direction $x$ (the calculation of these local
polarizations is described in Ref. \onlinecite{Wojde2013}). Black and blue points show the calculated polarization
components along $z$ and $y$, respectively, while lines are guide to the eye. A spontaneous $P_y$ polarization arises at the DWs in
addition to the monotonic inversion of $P_z$. \label{fig:supercell}}
\end{figure}

Little is known about the phonon spectral distribution in multidomain ferroelectrics. In particular, the mechanisms of phonon scattering at 
DWs have not been elucidated for any material to the best of our knowledge. Such type of information is difficult to extract from experiments and typically requires the 
combined use of theory and numerical simulations. In this regard, just a single theoretical study has very recently dealt with the problem 
of simulating phonon transport through multidomain ferroelectrics. \cite{Wang2016} However, a continuum-level phase-field approach was used in which 
the DW phonon scattering mechanism is a required input, rather than an output, of the calculation. 

Here we report for the first time phonon transport calculations in multidomain ferroelectrics with atomistic detail.
To this end, we interface force-constant matrices obtained from second-principles model potentials (SPMP) \cite{Wojde2013} with mode-dependent nonequilibrium Green's function
(NEGF) \cite{Ong2015} methodologies. SPMP is a powerful approach, recently developed by one of us, to study ferroelectric lattice-related properties 
with potential to provide a detailed atomistic picture of a temperature-driven DW formation,\cite{Wojde2014} as well as to predict
exotic properties such as negative capacitance in multidomain ferroelectric superlattices.\cite{Zubko2016} 
On the other hand, mode-dependent NEGF is an extension of the conventional NEGF method \cite{Sadasivam2014} to obtain 
the individual contribution of each phonon mode to the thermal transport. A thorough formulation of this 
method has been recently introduced in Ref. \onlinecite{Ong2015} and a specific description can be found in \cite{supmat}. As a brief overview, the physical system is partitioned along the heat transport
direction into a scattering region (channel) and two semi-infinite reservoirs (contacts), that are held at constant
temperatures, and the problem consists in calculating the ballistic transmission probability of each phonon from one contact to 
the other across the channel. In the calculation we take advantage of the system periodicity 
in the plane perpendicular to the heat transport: 
By Fourier-transforming we can decouple a 3D transport calculation into a set of 1D calculations with different 2D transverse wave 
vectors $\mathbf{k_{\perp}}$. The total measurable 3D magnitudes are then retrieved by integration over the 2D 
Brillouin zone.

An example of the atomistic computational supercell employed in the simulations is shown in Fig.~\ref{fig:supercell}. It
consists of a tetragonal $\mathrm{PbTiO_3}$ (PTO) lattice with the ferroelectric polarization oriented parallel to the $z$ 
axis. In the case of the figure, two flat 180$^{\circ}$ DWs have been introduced in the channel. The DWs lie in the $y,z$ plane, and are thus 
perpendicular to the transport direction $x$. As shown in panel Fig.~\ref{fig:supercell}(c),
across the leftmost DW the ferroelectric polarization $P_z$ transits monotonically from positive to negative values, and vice versa
across the rightmost DW. 
In addition, a spontaneous polarization along the $y$ direction ($P_y$) is formed at the DWs and rapidly vanishes into the domains. According
to Ref. \onlinecite{Wojde2014}, this polarization is switchable, so that parallel and antiparallel metastable 
configurations can in principle be induced in samples with more than one DW.

\begin{figure}
\includegraphics[width=0.8\linewidth]{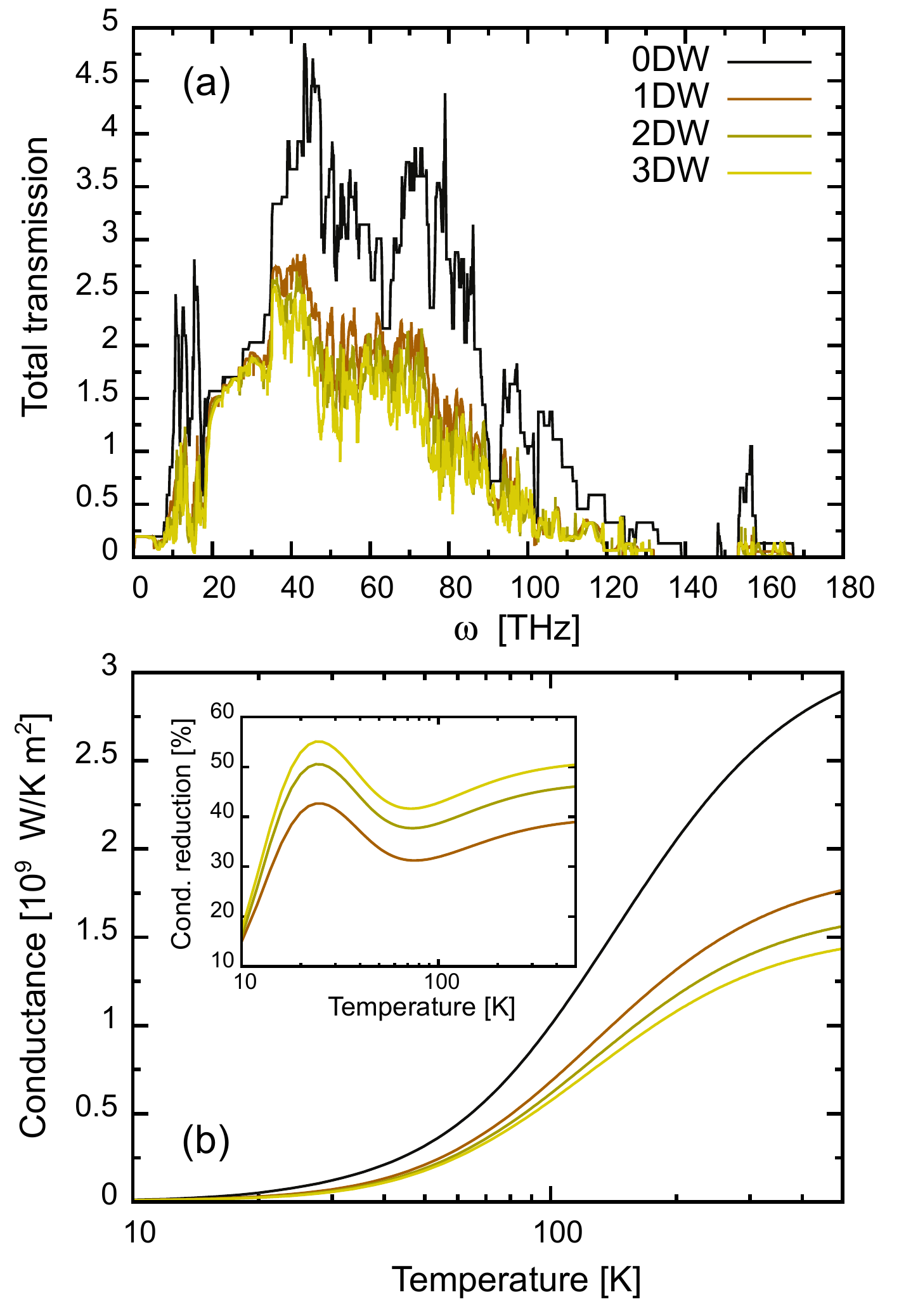}
\caption{(a) Total phonon transmission function and (b) Thermal conductance calculated for PTO samples with
0, 1, 2 and 3 flat DWs in the channel. The inset in panel (b) shows the \% of conductance reduction with respect to the 
monodomain (0DW) system. \label{fig:trans_cond}}
\end{figure}

We begin the heat transport study by analyzing the total phonon transmission function and thermal conductance calculated for PTO samples with up to three DWs in series in the
channel. The results are shown in Fig.~\ref{fig:trans_cond} together with those for a reference monodomain (0 DW) sample. It is observed that the 
presence of a single DW reduces considerably the phonon transmission. In particular, an efficient scattering is observed
for phonons of frequency higher than 35 THz. Interestingly, the DW does not work well as a low-pass phonon filter since a strong scattering is also
observed for three low-frequency peaks. The lattice thermal conductance is strongly depleted by a single DW with a maximal reduction of $\sim43$\% attained at $T \sim 25$ K and essentially related to the scattering of the low-frequency phonons. This is a remarkable effect, taking into account the coherent nature of the ideal DW we simulate. 

The inclusion of a second DW in the channel further reduces the transmission and conductance, but its effect is much smaller
than that of the first DW, a trend that is repeated for a third DW. Indeed, it seems that both transport magnitudes would 
tend towards an asymptotic limit if an increasing number of DWs in series were included in the channel. This indicates
that the DWs do not act on the thermal flux as series resistors; instead, most of the scattering is brought about by the first few DWs
and some phonons travel across the multiple interfaces without being scattered. Similar asymptotic behaviors have been observed in 
other type of multicomponent layered systems with few clean heterojunctions \cite{Abramson2002,Zhang2007,Tian2014} and in superlattices 
\cite{Luckyanova2012}. We have also checked that the DW spacing has a negligible effect on the transport properties in the current ballistic regime (see \cite{supmat}), 
although this is known to be a crucial factor when anharmonic Umklapp processes become the dominant 
scattering mechanism at higher temperatures.\cite{Ihlefeld2015,Hopkins2013}

\begin{figure}
\includegraphics[width=1.0\linewidth]{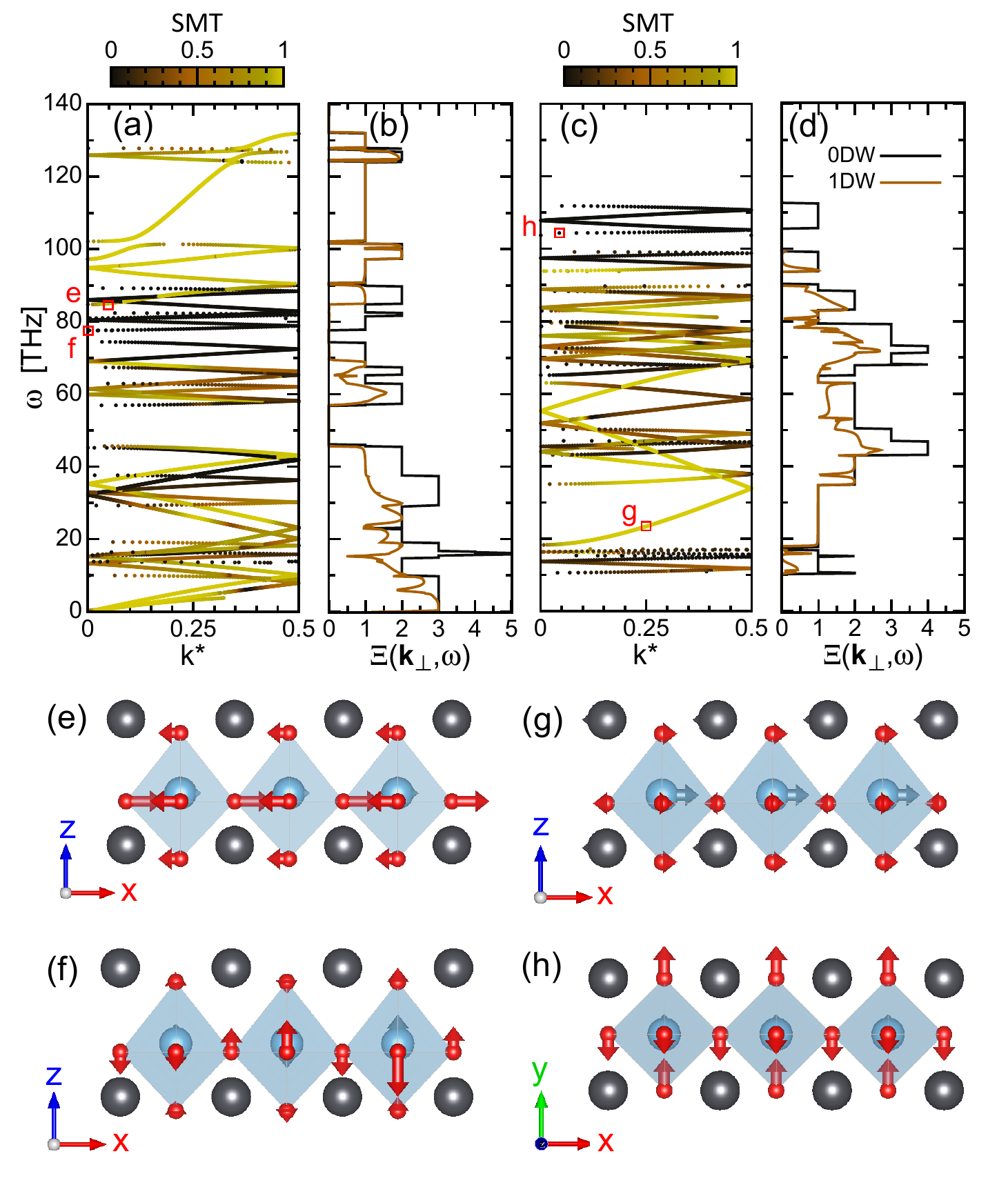}
\caption{Single-mode transmission analysis through a single DW shown for two selected values of $\mathbf{k}_{\perp}$. The dispersion
of the left contact phonons (the right contact bands are equivalent) along the longitudinal reduced wave vector is shown in panels
(a) and (c) for the transverse Brillouin zone centre ($k_y^*=k_z^*=0$) and zone edge ($k_y^*=k_z^*=0.5$), respectively. The bands are 
four times folded because four primitive unit cells have been used to define the contact principal layer along the longitudinal direction.
The colormap illustrates the individual probability of transmission of each phonon mode. The transmission functions 
at a discrete transverse wave vector ($\Xi(\mathbf{k}_{\perp},\omega$)) are obtained by summing over all phonon modes of
panels (a) and (c) at a given frequency and are respectively shown in (c) and (d) and compared with the transmission function of a monodomain sample. Notice
the integer values for the 0DW case that give the number of available phonon modes at a given ($\mathbf{k}_{\perp},\omega$) point.
Panels (e)-(h) show with arrows the atomic displacements of the $\Gamma$-phonon modes with highest contribution in the phonons indicated
with red squares in panels (a) and (c). Grey, blue and red colours are used for Pb, Ti and O atoms, respectively. \label{fig:SMT_analysis}}
\end{figure}

To elucidate the question of which phonons are being most effectively scattered by the DWs, we show in Fig.~\ref{fig:SMT_analysis}
the results of the single-mode transmission calculation at two discrete $\mathbf{k}_{\perp}$ points, namely, at the transverse Brillouin 
zone centre (panels (a) and (b)) and zone edge (panels (c) and (d)). The colormap in panels (a) and (c) illustrates the probability of
transmission of each individual phonon across a single DW and indicates that, as anticipated, some modes are completely scattered 
(black dots) while others are completely transmitted (light-yellow dots). One can even appreciate whole phonon bands with
a well-defined character, scattered or transmitted, occurring at different frequency ranges. We can get more insight into the nature
of the scattered and transmitted modes by analyzing their eigendisplacements. To this end we project the lattice-periodic part of the eigenvectors of selected
phonons, which in general have complex components, on the basis set formed from the phonon eigenvectors at the $~\Gamma$ point, which can 
be chosen to be real. We observe that the phonons scattered by the DW can be associated to $\Gamma$-modes whose corresponding atomic 
displacements are along the directions transverse to the heat flux, both parallel to the ferroelectric polarization
(see Fig.~\ref{fig:SMT_analysis} (f)) as well as perpendicular to it (see Fig.~\ref{fig:SMT_analysis} (h)). In contrast, transmitted
phonons modes are associated with atomic displacements along the longitudinal direction (see Figs.~\ref{fig:SMT_analysis} (e) and (g)).

\begin{figure}
\includegraphics[width=0.8\linewidth]{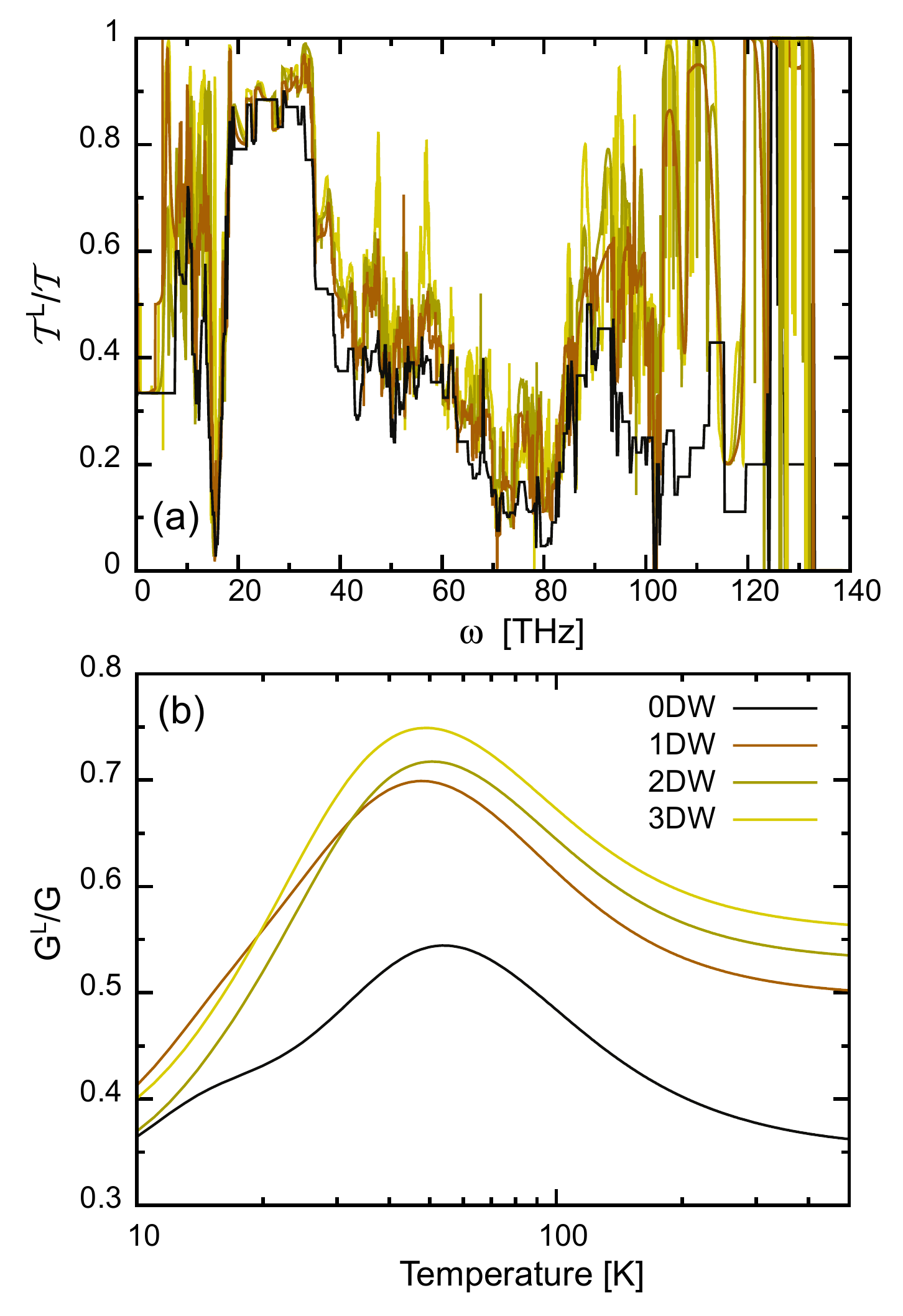}
\caption{ Fraction of total phonon transmission function (a) and thermal conductance (b) due to phonons that are at least 70\% 
polarized in the longitudinal direction for samples with different number of DWs in series.
\label{fig:polarizer_analysis_flat_DW}}
\end{figure}

Phonons cannot be classified as purely transverse or purely longitudinal vibrational modes except at special points of the Brillouin
zone. Here we adopt a practical approach and identify as longitudinal modes those phonons whose eigenvectors present at least 70\% polarization 
along the longitudinal $x$-direction. We then quantify the observed polarization-filter effect by computing the phonon 
transmission function ($\mathcal{T}^L$) and thermal conductance ($G^L$) due to the transport of longitudinal phonons.
In Figs.~\ref{fig:polarizer_analysis_flat_DW} (a) and (b) we show $\mathcal{T}^L$ and $G^L$ scaled by the total transmission function
and thermal conductance, respectively, in order to obtain comparable results for systems with different number of DWs. 
In other words, we represent the fraction of phonon transmission and thermal conductance due to longitudinal phonons.
Before commenting on the results for the samples with DWs, it is worth to observe that the thermal transport is  
dominated by longitudinal phonons even in the monodomain case (black curves). 
Nonetheless, the inclusion of a single DW in the channel increases considerably the contribution to the transport from longitudinal 
phonons, thereby polarizing the thermal flux that crosses the DW. A second and a third DW in series increase the polarizer effect though they do it in a gradually
smaller degree, approaching an asymptotic value in agreement with the results in Fig.~\ref{fig:trans_cond}. The maximal 
contribution from longitudinal phonons to the conductance ($\sim 75\%$), and consequently the most polarized thermal flux, is attained at $\sim 50$ K for the sample with 3 DWs. 
The polarizer effect is reduced at higher temperatures until it saturates at yet remarkable values. 
This demonstrates that series of 180$^{\circ}$ DWs in PTO work as longitudinal polarizers of high selectivity in lattice thermal transport.
We have also explicitly checked that the polarizer effect persists in DWs that are not ideally flat (see \cite{supmat} for our tests on this).

The present polarizer effect is clearly originated in the DW scattering of transverse phonons. 
A finer decomposition of the contribution to the heat transport from transverse modes parallel 
and perpendicular to the ferroelectric polarization demonstrates that 
both types are similarly scattered (see \cite{supmat}). Thus, bearing in mind the atomic deformations occurring
at the $180^{\circ}$ DWs we interpret that, first, the propagation of $z$-polarized phonons is impeded by the inversion of the ferroelectric 
distortion across the DWs. Second, since the atomic distortions along the $y$-direction are equivalent, i.e., null, at both sides of the DW, 
the $y$-polarized phonons must be scattered by the distortion associated to the occurrence of the $P_y$ polarization at the DW itself 
(see Fig.~\ref{fig:supercell}).\cite{Wojde2014} In this regard, we have observed that the parallel or antiparallel orientation of $P_y$ in multiple DW samples does not entail
any appreciable difference on the polarizer effect.

Following this reasoning, 180$^{\circ}$ PTO DWs are transparent for the propagation of longitudinal modes because the components
of the atomic distortions along the longitudinal $x$-direction are the same (in particular, null) at both domains and at the DW. 
Hence, it seems reasonable to predict the opposite effect, i.e., a transverse phonon polarizer, 
to occur for an alternative configuration of 180$^{\circ}$ DWs with ferroelectric interfaces perpendicular to the 
polarization. Such DWs, usually called head-to-head or tail-to-tail, have the peculiarity that the variation of electric displacement 
across them satisfies Maxwell's relation $\nabla D = \rho_{\rm free} \neq 0$. As a consequence, these walls can only occur in presence 
of free carriers or charged defects, and are difficult to obtain experimentally. \cite{Gureev2011,Bednyakov2015,Bednyakov2016} 
Yet, our results suggest that they may offer interesting opportunities for filtering effects. (Note that the walls investigated in 
this work satisfy $\nabla D = 0$. Studying charged DWs would require an explicit treatment of electronic effects in the simulations, 
and is beyond the scope of the lattice potentials used here.)

In conclusion, we have demonstrated that thermal transport across 180$^{\circ}$ DWs formed in bulk PTO is sensitive to the polarization
of the heat carrying phonons. The propagation of transverse phonons across the DWs is strongly suppressed while, in contrast, the walls are 
essentially transparent for the longitudinal modes. All in all 180$^{\circ}$ DWs behave as longitudinal phonon polarizers of high selectivity 
whose effect is robust against deviations from the ideal flat shape. Our results also suggest that other DW geometries 
--like e.g. the so-called head-to-head or tail-to-tail configurations-- may behave as transverse phonon polarizers. Thereby, multi-terminal 
devices capable to rewrite 180$^{\circ}$ DWs both parallel 
and perpendicular to the ferroelectric polarization can be envisaged as dynamically reconfigurable longitudinal/transverse polarizers. 
Finally, the concept has been here demonstrated for high-frequency heat carrying lattice vibrations due to the narrow width of the ferroelectric walls. 
Nevertheless, if it was possible to obtain wider DWs the same principle could apply for longer wavelength phonons eventually approaching the ultrasound transport regime.

\begin{acknowledgments}
M.R. and R.R. acknowledge financial support by the Ministerio de Econom\'ia y
Competitividad (MINECO) under grant FEDER-MAT2013-40581-P and by the
Severo Ochoa Centres of Excellence Program under Grant SEV-2015-0496
and by the Generalitat de Catalunya under grants no. 2014 SGR 301 and
through the Beatriu de Pin\'os fellowship program (2014 BP\_B 00101).
C.E.S and J.I. are funded by the Luxembourg National Research Fund
through the PEARL (Grant P12/4853155 COFERMAT), CORE (Grant
C15/MS/10458889 NEWALLS) and AFR (PhD Grant No. 9934186 for C.E.S.)
programs.
We thank the Centro de Supercomputaci\'on de Galicia (CESGA) for the 
use of their computational resources.
\end{acknowledgments}



\bibliography{./library}

\end{document}